\documentclass[runningheads]{svmult}

\usepackage{graphicx}
\usepackage{physprbb}
\usepackage{amssymb}  

\newcommand{\greeksym}[1]{{\usefont{U}{psy}{m}{n}#1}}

\newcommand{\udelta}{\mbox{\greeksym{d}}}
\newcommand{\uDelta}{\mbox{\greeksym{D}}}

\begin{document}

\title*{Aperiodicity and Disorder -- Does it Matter?}

\toctitle{Aperiodicity and Disorder -- Does it Matter?}

\titlerunning{Aperiodicity and Disorder}

\author{Uwe Grimm}

\authorrunning{Uwe Grimm}

\institute{Applied Mathematics Department, 
           Faculty of Mathematics and Computing,\\
           The Open University,
           Milton Keynes MK7 6AA, U.K.}

\maketitle

\begin{abstract}
The effects of an aperiodic order or a random disorder on phase
transitions in statistical mechanics are discussed. A heuristic
relevance criterion based on scaling arguments as well as specific
results for Ising models with random disorder or certain kinds of
aperiodic order are reviewed. In particular, this includes an exact
real-space renormalization treatment of the Ising quantum chains with
coupling constants modulated according to substitution sequences,
related to a two-dimensional classical Ising model with layered
disorder. 
\end{abstract}

\section{Introduction}

Equilibrium statistical mechanics bridges the gap between the
fundamental laws of physics at microscopic scales on the one hand and
the macroscopic phenomena that we may experience in every-day life on
the other hand \cite{Bal}. Its main achievement is the understanding
of the behaviour of macroscopic systems on the basis of the
microscopic constituents and their mutual interactions.

One of the most fascinating topics in this field concerns phase
transitions, i.e., a qualitative change observed in some physical
properties of a macroscopic system after a slight variation of certain
parameters as, for instance, the temperature or an external magnetic
field. Here, statistical mechanics reveals its full power by being
able to explain how even short-range interactions at the microscopic
scale may lead to such violent cooperative behaviour.  In mathematical
terms, a phase transition corresponds to a non-analyticity of the
thermodynamic potential that describes the equilibrium state of the
system as a function of the external parameters. In essence, one may
regard it as a battle between energetic and entropic terms, the first
driving the system towards an ordered phase at low temperature, the
second favouring a disordered state at higher temperatures. However,
this is a somewhat simplistic view of a very complex phenomenon that
should not be overstressed. As an example that teaches one to be
cautious one might think of the entropy-driven phase transitions
observed in non-interacting hard-core systems \cite{Fre} that,
counter-intuitively, may lead to an ordered state that has a higher
entropy than the disordered state.

While the general picture is rather well understood, it turns out to
be much more difficult to calculate the properties of a specific
macroscopic system from its microscopic interactions explicitly. In
fact, only for a few simple models complete solutions are known
\cite{Bax}. Among these is the paradigm of a ferromagnet, the
celebrated Ising model \cite{Lenz,Ising}, after Onsager \cite{Ons}
accomplished to calculate the free energy for the two-dimensional
Ising model without an external magnetic field.  However, also this
extremely simplistic model has, so far, defied any attempts at an
analytic solution in higher dimensions, or even in two dimensions in
the presence of an external magnetic field. Recent progress from
complexity theory might even support the widespread belief of its
general unsolvability.

Moreover, the models that have been solved analytically (or `exactly',
which usually does not mean rigorously in the mathematical sense) are
almost exclusively tied to regular periodic structures, as for
instance the Ising model on a square lattice or some other regular
planar lattices.  However, Nature is never completely regular, and
this immediately spurs the question what happens if the regular
periodic structure is disturbed by some randomness, mimicking a more
realistic crystalline substance, or if it even is given up completely,
resulting in either aperiodically ordered, or disordered
structures. While the effects of a random disorder have been studied
for many decades, aperiodically ordered systems entered the scene
after the experimental discovery of aperiodically ordered solids known
as \emph{quasicrystals} in the 1980s, see \cite{SSH} for a recent
compilation of introductory lectures on this issue.

The situation is, however, not as hopeless as it may seem on first
view. One of the things that come at our rescue is the important
concept of \emph{universality} of critical phenomena
\cite{Kada}. According to the universality hypothesis, the critical
behaviour of a system does \emph{not} depend on details of the
microscopic interactions of the model, but merely on a number of
general properties such as the dimension of space and the symmetries
of the model, see also the contributions \cite{Binder,RAR,Schwabl,TV}
in this volume. This means that second-order or `smooth' phase
transitions, which are characterized by an infinite correlation
length, can be classified into \emph{universality classes} of systems
showing the same physical behaviour at and close to the critical
point. In particular, this implies that the critical exponents that
describe the singularities of physical quantities upon approaching the
critical point only depend on the universality class of a model. The
origin of universality and the \emph{scaling relations} between
different critical exponents can be understood in terms of the
so-called \emph{renormalization group} \cite{Car,Binder} because
critical points correspond to fixed points under the renormalization
flow.  Of course, in reality things are not quite as simple, and the
situation may, in general, turn out to be more complex. Therefore,
such arguments should be taken with a grain of salt.

In two dimensions, we have another powerful tool at our disposal.
There exists an intimate relation between two-dimensional statistical
models at criticality and \emph{conformal field theory} \cite{DFMS}
that explains the surprising fact that critical exponents in many
cases are found to be rational numbers, and furthermore allows the
computation of correlation functions. This yields an alternative, and
more substantiated, explanation of the universality classes and a
partial classification of the possible types of critical behaviour in
two dimensions. Such strong, albeit by no means complete, results are
limited to the two-dimensional case due to the particular properties
of the two-dimensional conformal symmetry group.

Nevertheless, because of universality, one may expect a certain
robustness of critical point properties, i.e., introducing disorder or
aperiodic order may not affect the critical behaviour of some systems
at all. The obvious question one would like to answer is, given an
aperiodic or disordered model, does it show the same critical
behaviour as its perfectly ordered fellow or not? And, if not, how
does it change?

In what follows, we shall concentrate on the example of the Ising
model, simply because it is the most frequently studied and, besides
numerous numerical treatments, also permits analytic solutions.  After
a brief introduction to the Ising model and its critical properties in
the subsequent section, we discuss a rather general, albeit heuristic,
relevance criterion that is based on scaling arguments in
section~\ref{sec:harrisluck}. Thereafter, in
section~\ref{sec:disorder}, we give a short overview on the various
types of disordered Ising models considered in the literature and the
main results concerning their critical behaviour. Next we turn our
attention to aperiodically ordered Ising models which are the topic of
section~\ref{sec:aperiodic}. Finally, the article concludes with a
brief summary in section~\ref{sec:conc}.

\section{The Ising Model}
\label{sec:Ising}

There have been numerous review articles and even entire books
\cite{McCoyWu} devoted to the Ising model, and many of these give en
extensive introduction into the history of this simple, yet
fascinating model. Also most textbooks on statistical mechanics treat
at least the one-dimensional Ising model. For a mathematically
rigorous approach, see, e.g., \cite{Georgii}. Here, we shall restrict
ourselves to the most important facts, the interested reader is
referred to the literature for a more complete account of the
historical background, for instance in \cite{Kobe}. Also the recent
review \cite{GB} on aperiodic Ising models contains a brief historical
intreoduction, and an extensive list of references.

\subsection{The Square-Lattice Ising Model}

Originally, the Ising model was devised by Lenz as a simplistic toy
model of a ferromagnet \cite{Lenz}. The one-dimensional model was
later solved by Ising \cite{Ising}, a student of Lenz, who obtained
the disappointing result that the model does not show a phase
transition at a non-zero temperature. He supposed that this should
hold in higher dimensions as well, but, fortunately, it did not take
too long until he was proven wrong -- otherwise the model carrying his
name would not have attracted the interest of physicists for almost a
century.

\begin{figure}[tb]
\begin{center}
\includegraphics[width=0.80\textwidth]{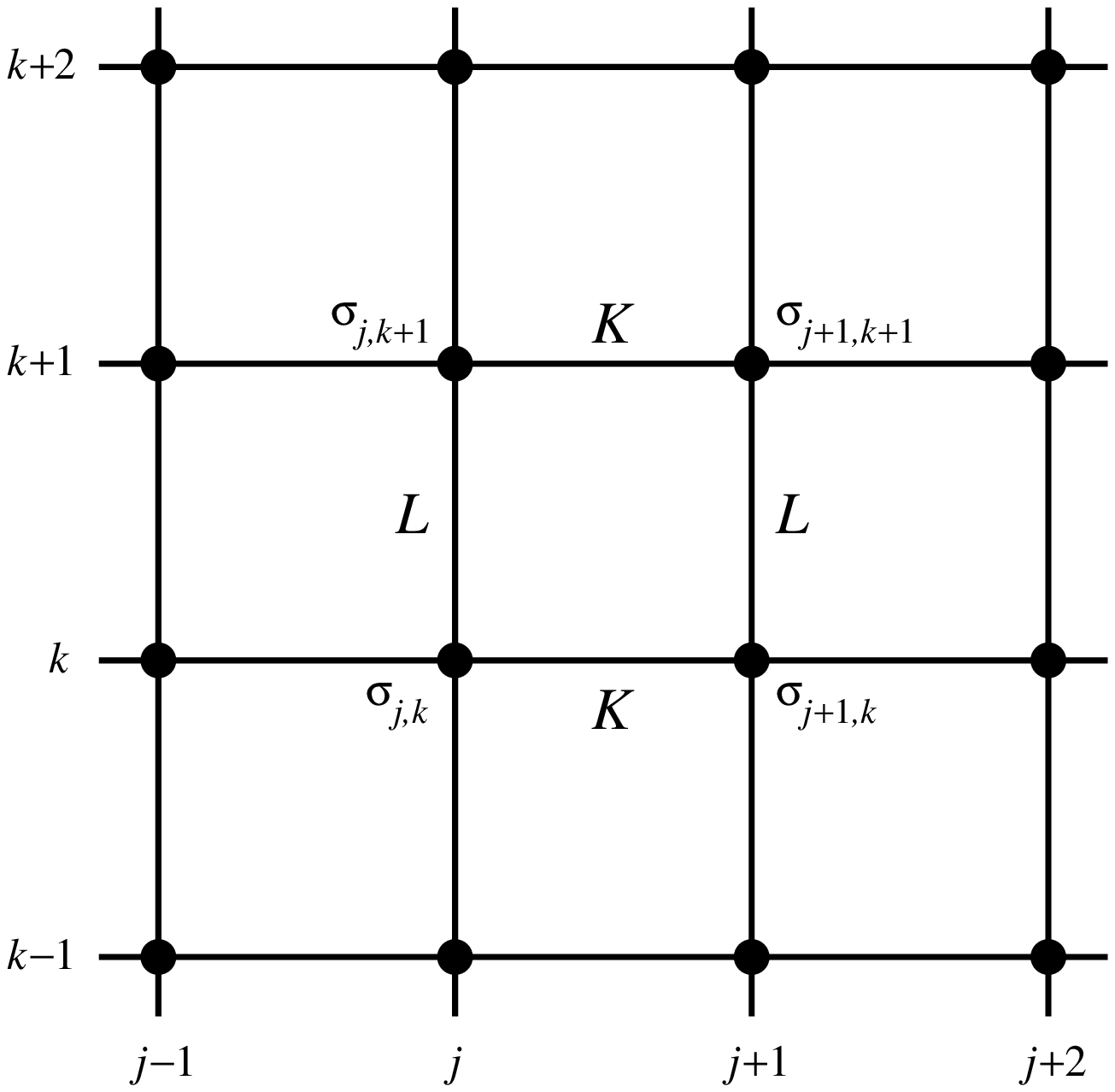}
\end{center}
\caption[]{The square-lattice Ising model}
\label{fig:sq}
\end{figure}

So, how is the model defined? The local magnetic moments of the solid
are modeled by `Ising spins' on the sites of a regular lattice, which
can only take two values, $\pm 1$. The ferromagnetic interaction
between these spins is given by a simple coupling of neighbours on the
lattice. Let us concentrate on the square-lattice case for
definiteness. A \emph{configuration}
$\sigma=(\sigma_{1,1}^{},\sigma_{1,2}^{},\ldots,\sigma_{N,M}^{})\in\{\pm
1\}^{NM}$ of spins $\sigma_{j,k}^{}$ on a $N\times M$ rectangular
section of the lattice is assigned an energy
\begin{equation}
E(\sigma) = -\sum_{j=1}^{N}\sum_{k=1}^{M} \left(
K\sigma_{j,k}^{}\sigma_{j+1,k}^{} + 
L\sigma_{j,k}^{}\sigma_{j,k+1}^{} + 
H\sigma_{j,k}^{} \right)
\label{eq:im}
\end{equation}
where we may assume certain boundary conditions such as, for instance,
periodic ($\sigma_{j,M+1}^{}\equiv\sigma_{j,1}^{}$,
$\sigma_{N+1,k}^{}\equiv\sigma_{1,k}^{}$) or free
($\sigma_{j,M+1}^{}\equiv\sigma_{N+1,k}^{}\equiv 0$) boundary
conditions. Here, $K>0$ and $L>0$ denote the ferromagnetic coupling
constants in horizontal and vertical direction, respectively, see
Fig.~\ref{fig:sq}, and $H$ is the magnetic field.

The \emph{partition function}, a sum of Boltzmann factors over all
configurations $\sigma$, is given by
\begin{equation}
Z_{N,M}(T,H) = \sum_{\sigma}\exp[-E(\sigma)/k_{\mathrm{B}}^{}T]
\end{equation}
where $T$ denotes the temperature and $k_{\mathrm{B}}^{}$ is
Boltzmann's constant.  It can also be expressed in terms of a
so-called \emph{transfer matrix}
\begin{equation}
Z_{N,M}(T,H) = \mbox{Tr}[U_{N}^{}(T,H)^M]
\label{eq:tm}
\end{equation}
where $U_{N}^{}$ is a $2^{N}_{}\times 2^{N}_{}$ matrix whose elements
are Boltzmann weights of the configuration corresponding to a single
row of the lattice, and the trace is due to the periodic boundary
conditions that we assumed. Thus, the trace performs the sum over the
configuration
$(\sigma_{1,1}^{},\sigma_{2,1}^{},\ldots,\sigma_{N,1}^{})$ of the
first row, whereas the summation on the configurations
$(\sigma_{1,k}^{},\sigma_{2,k}^{},\ldots,\sigma_{N,k}^{})$, $2\le k\le
N$, along the other rows are hidden in the matrix product.

\subsection{Phase Transitions and Critical Exponents}

In the thermodynamic limit of an infinite system, the \emph{free
energy} per site
\begin{equation}
f(T,H) = \lim_{N,M\rightarrow\infty}\left(\frac{-k_{\mathrm{B}}^{}T}{NM}
\ln [Z_{N,M}(T,H)]\right)
\end{equation}
is an analytic function of the parameters $T$ and $H$, except at the
phase transition. The transition is of first order if already a first
derivative of the free energy is discontinuous.  Otherwise, the
transition is termed a second- or higher-order transition. Examples of
first-order transitions are melting of ice or the evaporation of
water, where one has to supply energy, the latent heat, to the system
without changing the temperature. The Curie point of a ferromagnet, or
the critical point of water at the end of the first-order transition
line between liquid and gas, are examples of critical points.

Correspondingly, one encounters singularities in thermodynamic
quantities that can be obtained as derivatives of the free energy such
as the specific heat
\begin{equation}
c_{H}^{}(T) = - T \frac{\partial^{2} f(T,H)}{\partial T^{2}_{}}
\end{equation}
at constant magnetic field $H$, the spontaneous magnetization
\begin{equation}
m_{0}^{}(T)=\lim_{H\searrow 0}m(T,H) \; ,\qquad
m(T,H)=-\frac{\partial f(T,H)}{\partial H} \; ,
\end{equation}
or the magnetic susceptibility
\begin{equation}
\chi_{T}^{}(H)=\frac{\partial m(T,H)}{\partial H}=
-\frac{\partial^2 f(T,H)}{\partial H^2}
\end{equation}
For a second-order phase transition, on which we shall now
concentrate, these singularities are characterized by so-called
\emph{critical exponents} describing power-law singularities on
approaching the critical point.  For magnetic systems such as the
Ising model, some zero-field ($H=0$) critical exponents are
\begin{eqnarray}
c_0^{}(T) &\sim& |t|^{-\alpha}\\
m_0^{}(T) &\sim& (-t)^{\beta} \qquad \mbox{(for $t<0$, $m_0^{}(T)\equiv 
0$ for $t\ge 0$)}\\
\chi_{T}^{}(0)   &\sim& |t|^{-\gamma}
\end{eqnarray}
as $t\rightarrow 0$, where $t=(T-T_{\mathrm{c}}^{})/T_{\mathrm{c}}^{}$
is a `reduced temperature' parametrizing the distance to the critical
temperature $T=T_{\mathrm{c}}^{}$, see also \cite{TV}. The power-laws
are related to a particular behaviour of correlation $G(r,r')$ at the
critical point. Instead of a usual exponential decay with the
distance, $G(r,r')\sim\exp(-|r-r'|/\xi)$, with a characteristic
correlation length $\xi$, correlation functions at a critical point
show a power-law decay $G(r,r')\sim|r-r'|^{-x})$. In other words, the
correlation length $\xi$ diverges as
\begin{equation}
\xi\sim |t|^{-\nu}
\end{equation}
as the critical point is approached. Of course, several other
exponents can be considered, and there also exist phase transitions
where the exponents differ when the critical point is approached from
the ordered phase $t<0$ or from the disordered phase $t>0$.

\subsection{The Ising Model Phase Transition}

But as mentioned above, it was not clear from the beginning that the
Ising model has a phase transition at all. However, already in 1936,
Peierls \cite{Peierls} gave an argument that showed the existence of a
phase transition in two or more dimensions. The argument is simple,
and indeed also holds for rather general models, including Ising
models with ferromagnetic random couplings.

Essentially, the argument goes as follows. At $T=0$, the system is in
one of its magnetized ground states, where all spins are either $+1$
or $-1$. The states that the system can explore at a small temperature
differ from the ground state by small islands of turned spins, and
their energy is given by the surface of these islands where
neighbouring spins differ. So, even at small finite temperatures, the
system is magnetized, and there must be a transition to a disordered
high-temperature state at a non-zero temperature. Indeed, this
argument breaks done in one dimension, because a flip of the local
spins along an entire infinite half-line affects a single bond only,
thus even an arbitrarily small temperature suffices to destroy the
magnetic order, see \cite{Ellis} for a more precise discussion.

\begin{figure}[b]
\begin{center}
\includegraphics[width=0.81\textwidth]{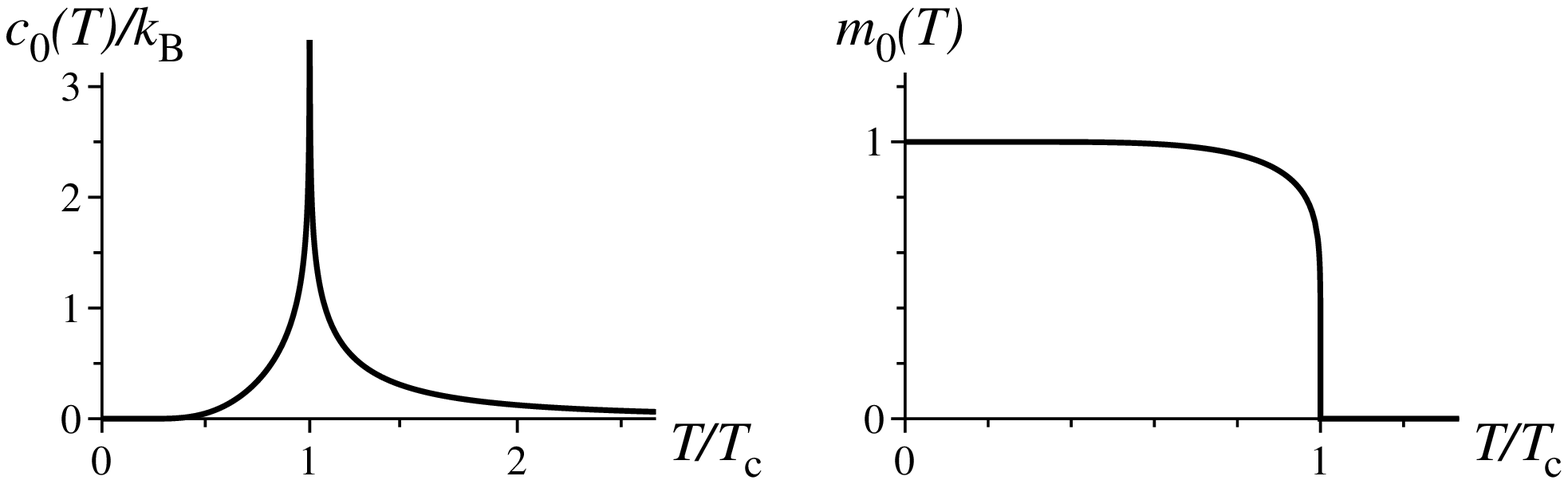}
\end{center}
\caption[]{The specific heat $c_{0}^{}(T)$ and the spontaneous 
magnetization $m_{0}^{}(T)$ of the square-lattice Ising model}
\label{fig:crit}
\end{figure}

The critical temperature for the zero-field Ising model was first
determined by Kramers and Wannier \cite{KW} in 1941 under the
assumption that there is only a single phase transition. In this case,
the transition has to take place at a fixed point of a duality
transformation that maps low- and high-temperature phases onto each
other. For the square-lattice Ising model, the critical temperature
$T_{\mathrm{c}}^{}$ is given by
\begin{equation}
\sinh(2K/k_{\mathrm{B}}{}T_{\mathrm{c}}{})
\sinh(2L/k_{\mathrm{B}}{}T_{\mathrm{c}}{}) = 1\; ,
\qquad
H=0\; ,
\end{equation}
Only three years later, Onsager's celebrated solution of the
zero-field square-lattice Ising model \cite{Ons} revealed a
logarithmic singularity of the specific heat,
\begin{equation}
c_{0}{}(t)\sim \ln |1/t| \; ,
\label{eq:sh}
\end{equation}
i.e., the corresponding critical exponent $\alpha=0$. The correlation
length diverges at the critical point with an exponent $\nu=1$. The
magnetization exponent $\beta=1/8$ was later calculated by Yang
\cite{Yang} by means of a perturbative approach. In
Fig.~\ref{fig:crit}, the specific heat and the spontaneous
magnetization for the zero-field Ising model are shown.

\subsection{The Ising Quantum Chain}

Before we can, finally, move on to discuss disordered models, we need
to introduce a one-dimensional quantum version of the Ising model
\cite{St} that is closely related to the two-dimensional classical
Ising model, see also \cite{TV}. Consider the Hamiltonian
\begin{equation}
H =-\frac{1}{2}
\left(\sum\limits_{j=1}^{N}
     \sigma^{x}_{j}\sigma^{x}_{j+1} +
     \lambda\sum\limits_{j=1}^{N} \sigma^{z}_{j}\right)
\label{eq:ham}
\end{equation}
on the Hilbert space
$\bigotimes_{j=1}^{N}\mathbb{C}^{2}_{}\cong\mathbb{C}^{2^N}_{}$. Here,
the local spin operators are defined as
\begin{equation}
\sigma^{x,z}_{j}=
\underbrace{\mathbb{I}\otimes\mathbb{I}\otimes\ldots\otimes
\mathbb{I}}_{\mbox{\scriptsize $j\! -\! 1$ factors}}
\otimes\sigma^{x,z}_{}\otimes
\underbrace{\mathbb{I}\otimes\ldots\otimes\mathbb{I}\otimes
\mathbb{I}}_{\mbox{\scriptsize $N\! -\! j$ factors}}
\end{equation}
with the standard Pauli matrices
\begin{equation}
\sigma^{x}_{}=\left(\begin{array}{@{\,}r@{\;\;}r@{\,}}
0&1\\ 1&0\end{array}\right)\; ,
\qquad
\sigma^{z}_{}=\left(\begin{array}{@{\,}r@{\;}r@{\,}}
1&0\\ 0&-1\end{array}\right)\; ,
\qquad
\mathbb{I} =\left(\begin{array}{@{\,}r@{\;\;}r@{\,}}
1&0\\ 0&1\end{array}\right)\; .
\end{equation}
Frequently, a different basis is used, where the coupling term
involves the products $\sigma^{z}_{j}\sigma^{z}_{j+1}$, and the
`transversal field' term has the form $\sum_{j=1}^{N}\sigma^{x}_{j}$
\cite{TV}. For the homogeneous Hamiltonian (\ref{eq:ham}), the
spectrum is known completely \cite{LSM}.

The Hamiltonian (\ref{eq:ham}) can be obtained as an anisotropic limit
of the transfer matrix $U_N^{}$ (\ref{eq:tm}) of the zero-field
($H=0$) square-lattice Ising model (\ref{eq:im}), where the coupling
along the chain direction $K\!\rightarrow\! 0$ and the coupling in the
`time' direction $L\!\rightarrow\!\infty$, while $\beta K\exp(2\beta
L)$ is kept fixed \cite{Kogut}. Consequently, the `transversal field'
parameter $\lambda$ in the quantum chain acts analogously to the
temperature in the classical model; and the Ising quantum chain at
zero temperature displays a quantum phase transition \cite{TV}, i.e.,
a change in the ground-state properties, at the critical value
$\lambda_{\mathrm{c}}^{}=1$ which belongs to the same universality
class as the critical point of the classical square-lattice Ising
model. In particular, the ground-state energy per site, $-E_{0}^{}/N$,
of the Hamiltonian (\ref{eq:ham}) corresponds to the free energy of
the classical Ising model, and the energy gap $E_{1}^{}-E_{0}^{}$
between the ground state and the first excited state corresponds to
the inverse $\xi_{\parallel}^{-1}$ of the correlation length along the
chain, i.e., the energy gap vanishes at criticality.  According to
finite-size scaling \cite{Binder,Her}, the gap vanishes as
$E_{1}^{}-E_{0}^{}\sim N^{-z}$, where
$z=\nu_{\parallel}^{}/\nu_{\perp}^{}$ denotes the ratio of the
correlation length exponents along the chain and in the `time'
direction. In spite of the anisotropic limit, at criticality the
homogeneous system (\ref{eq:ham}) behaves isotropically, with
$\nu_{\parallel}^{}=\nu_{\perp}^{}=1$ as in the classical
square-lattice Ising model. However, as we shall see below, this may
be different if one considers disordered or aperiodically ordered
Ising quantum chains, which then correspond to classical
two-dimensional Ising models with a layered disorder or aperiodicity.

\subsection{Effects of Disorder: Some General Remarks}

Phase transitions are cooperative phenomena leading to singularities
in physical quantities such as the specific heat or the magnetic
susceptibility. {}From a na\"{\i}ve point of view, any disorder in a
system, if it affects the phase transition at all, will tend to
\emph{weaken} critical singularities -- it is hard to imagine how it
could possibly do the opposite.  Thus, one might expect that a
first-order transition in a perfectly ordered system may be weakened
to a higher-order phase transition in a disordered or an aperiodically
ordered system, and indeed there exist examples where such behaviour
has been observed \cite{CBB}. For a second-order transition, disorder
may change the critical exponents, or the singularities may be
weakened to higher-order singularities, or even washed out completely
so that the phase transition disappears.

So, given a specific model, how can one estimate the effect of a certain
type of disorder on the system? A quite simple answer to
this question is provided by a heuristic relevance criterion based on
scaling arguments. Although these arguments rely on several
assumptions that may not always be fulfilled, they have proven to be
rather successful in predicting the correct critical behaviour.

\section{Heuristic Scaling Arguments}
\label{sec:harrisluck}

The relevance criterion that we are going to discuss now had first
been put forward be Harris in 1974 \cite{Har} for disordered Ising
models. Later, it was generalized by Luck \cite{Luck} to the case of
aperiodically ordered Ising models. The argumentation presented here
closely follows the discussion of the Harris-Luck criterion in
\cite{Her}, where one can also find further applications of the
criterion.

\begin{figure}[tb]
\begin{center}
\includegraphics[width=0.68\textwidth]{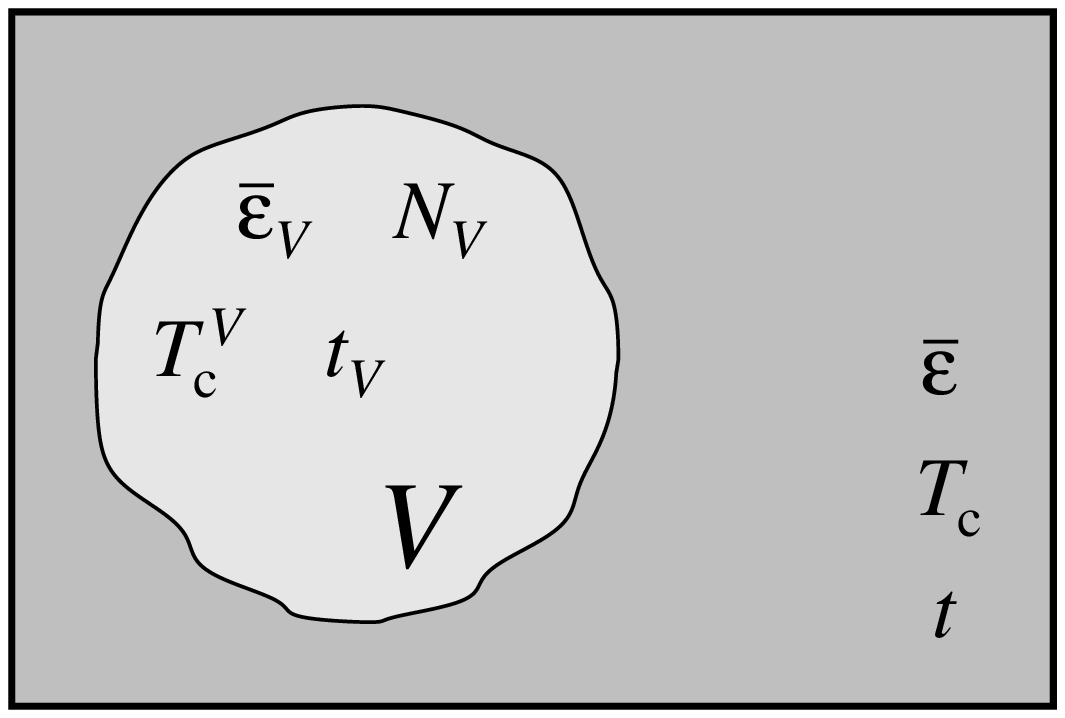}
\end{center}
\caption[]{A volume $V$ within the 
system and the associated local quantities}
\label{fig:hl}
\end{figure}

We restrict ourselves to models of Ising type with purely
ferromagnetic couplings; the case of frustration in disordered systems
or spin glasses, see \cite{Engel}, is out of reach and thus not
discussed here. Furthermore, a certain homogeneity of the distribution
of the ferromagnetic coupling constants $\varepsilon_{j,k}^{}\ge 0$ is
needed, such that, for instance, the mean coupling
$\bar{\varepsilon}=\varepsilon_{0}^{}$ is well defined, which we shall
consider as the coupling of our unperturbed reference system denoted
by a lower index $0$. We characterize the distribution of coupling
constants by their \emph{fluctuations}, expressed in terms of the
deviations from the mean coupling. More precisely, consider an
approximately spherical volume $V$ located somewhere in our infinite
system, see Fig.~\ref{fig:hl}. Within $V$, the mean coupling is
\begin{equation}
\bar{\varepsilon}_{V}^{} = 
\frac{2}{n_{0}^{}N_{V}^{}}
\sum_{\langle j,k\rangle\in V} \varepsilon_{j,k}^{}
\end{equation}
where $N_{V}^{}$ denotes the number of spins in $V$, and $n_{0}^{}$ is
the mean coordination number, i.e., the averaged number of neighbours
in the system.  The deviation of the accumulated couplings from the
mean is given by
\begin{equation}
G_{V}^{} \; =\; \sum_{\langle j,k\rangle\in V}\!
\varepsilon_{j,k}^{} - \frac{\varepsilon_{0}^{}n_{0}^{}N_{V}^{}}{2}
\; =\; \frac{n_{0}^{}N_{V}^{}}{2}
\left(\bar{\varepsilon}_{V}^{}-\varepsilon_{0}^{}\right)\; .
\end{equation}
As a measure of the fluctuations, we consider the asymptotic behaviour
of the standard deviation $\uDelta_{G_{V}^{}}^{}$ for large volumes
$V$,
\begin{equation}
\uDelta_{G_{V}^{}}^{} \sim (N_{V})^{\omega}_{}\; ,
\label{eq:omega}
\end{equation}
thus defining the \emph{fluctuation exponent} $\omega$ which we assume
to be well defined. Note that we did not specify the type of disorder
in the system, we may consider both models with varying coupling
constants on a regular lattice, or models that are defined on
topologically disordered graphs, for instance random graphs or
periodic lattices with randomly distributed vacancies. Another example
we shall encounter below concerns Ising models that are defined on
aperiodic graphs, which also yield fluctuations albeit there is no
randomness involved.

Now comes the crucial intuitive step. Clearly, the critical
temperature $T_{\mathrm{c}}^{}$ depends on the average coupling
constant $\bar{\varepsilon}=\varepsilon_{0}^{}$. Thus, due to the
fluctuations in the couplings $\varepsilon_{j,k}^{}$, the system may
\emph{locally}, i.e., within a volume $V$, correspond to a different
`local critical temperature' $T_{\mathrm{c}}^{V}$ that, for
sufficiently small disorder, may be expected to depend linearly on the
local average coupling
$T_{\mathrm{c}}^{V}\sim\bar{\varepsilon}_{V}^{}$. One may think of
$T_{\mathrm{c}}^{V}$ as the critical temperature of a system that
essentially looks the same in any ball of volume $V$. Correspondingly,
we may define a `local reduced temperature' $t_{V}^{}$.

If the disorder is \emph{irrelevant} for the critical behaviour, the
shift in the local reduced temperature $\udelta t=t_{V}^{}-t$ has to
vanish at criticality, i.e., $\udelta t\rightarrow 0$ as $t\rightarrow
0$. The relevance criterion now follows from this consistency
requirement by using scaling relations involving the critical exponent
$\nu$, which in the case of \emph{irrelevant} disorder should coincide
with the exponent $\nu_{0}^{}$ of the pure system. The argument
proceeds as follows. The volume $V$ of spins that are correlated is of
the order $V\sim\xi^{d}$, where $\xi$ denotes the correlation length,
and $d$ the dimension of the system. Then, from (\ref{eq:omega}) and
our assumption $T_{\mathrm{c}}^{V}\sim\bar{\varepsilon}_{V}^{}$, we
deduce
\begin{equation}
\udelta t\sim \xi^{-d(1-\omega)}\sim |t|^{\nu d(1-\omega)}\; ,
\end{equation}
where $\nu$ is the correlation exponent of the disordered system.
In the irrelevant scenario $\nu=\nu_{0}^{}$, this yields
\begin{equation}
\udelta t/t \sim |t|^{-\Phi} \qquad 
\mbox{with $\Phi=1-d\nu_{0}^{}(1-\omega)$}\; .
\end{equation}
Consequently, the disorder is \emph{irrelevant} if the crossover
exponent $\Phi<0$. 

We can summarize the Harris-Luck relevance criterion as follows
\cite{Har,Luck,Her}. A modulation in $d_{\mathrm{m}}^{}\le d$
space dimensions of a ferromagnetic $d$-dimensional Ising model
with correlation exponent $\nu_{0}^{}$ is \emph{relevant} for
$\omega>\omega_{\mathrm{c}}^{}$, \emph{marginal} for
$\omega=\omega_{\mathrm{c}}^{}$, and \emph{irrelevant} for
$\omega<\omega_{\mathrm{c}}^{}$, where
\begin{equation}
\omega_{\mathrm{c}}^{} = 1-\frac{1}{d_{\mathrm{m}}\nu_{0}^{}}
\end{equation}
and $\omega$ denotes the fluctuation exponent (\ref{eq:omega}).

For the two-dimensional Ising model, we have $\nu_{0}^{}=1$. Thus, for
a model with layered disorder, i.e., $d_{\mathrm{m}}^{}=1$, we obtain
$\omega_{\mathrm{c}}^{}=0$. In this case, any divergent fluctuation is
relevant. For planar disorder, i.e., $d_{\mathrm{m}}^{}=2$, the
marginal value is $\omega_{\mathrm{c}}^{}=1/2$. Thus, for randomly
distributed couplings, which correspond to $\omega=1/2$, one is
precisely in the marginal situation where the criterion does not give
a definite prediction. Generally, one might expect to find logarithmic
corrections to scaling in the marginal case, and this is indeed the
case in the two-dimensional random-bond Ising model
\cite{DD,Shalaev,Shankar,Ludwig}.

In the original formulation by Harris, the criterion was expressed in
terms of the specific heat exponent $\alpha$ in place of the
correlation exponent $\nu$.  The two exponents are related by the
hyperscaling relation $d\nu=2-\alpha$ \cite{Bax}, which holds for the
two-dimensional Ising model, which again corresponds to the marginal
case $\alpha=0$ in the case of random disorder \cite{Har}.

\section{Random Ising Models}
\label{sec:disorder}

Traditionally, randomness was introduced into Ising models to describe
magnetic behaviour of alloys where some magnetic atoms were replaced
by non-magnetic atoms. Therefore, such models are termed \emph{dilute}
Ising models, and, in the simplest case, just correspond to a regular
Ising model where some spins, or some bonds, have been removed, see
the reviews \cite{Stinch,Selke,SST}. Clearly, there is some relation
to percolation \cite{RAR}, because one at least needs infinite
clusters in order to support a phase transition. Somewhat more general
are random-bond Ising models, where the coupling constants are chosen
randomly, often from a bimodal distribution.

In these systems, the disorder is regarded as `frozen' or `quenched',
which means that it is static. Therefore, one has to average on the
free energy on the possible realizations, which is extremely difficult
\cite{Engel}. Nevertheless, some analytical results have been
obtained. As shown above, we expect the two-dimensional system to be
marginal with respect to random disorder. Using the replica method,
Dotsenko and Dotsenko \cite{DD} found a double logarithmic singularity
in the specific heat
\begin{equation}
c(t)\sim\ln[1+g\ln(1/|t|)] \; ,
\label{eq:dissh}
\end{equation}
in place of the pure Ising behaviour (\ref{eq:sh}), and
\begin{equation}
m(t) \sim \exp\{-a[\ln\ln(1/|t|)]^2/2\}\; ,\qquad 
\chi(t) \sim t^{-2}\exp\{-a[\ln\ln(1/|t|)]^2\}\; ,
\label{eq:DD}
\end{equation}
corresponding to critical exponents $\beta=0$ and $\gamma=2$, which
differ from the pure Ising values.  However, by means of bosonization
techniques and conformal field theory, Shalaev, Shankar, and Ludwig
\cite{Shalaev,Shankar,Ludwig} arrived at different results
\begin{equation}
m(t) \sim t^{1/8}(\ln |t|)^{-1/16}\; ,\qquad
\chi(t) \sim t^{-7/4}(\ln|t|)^{7/8}\; ,
\label{eq:SSL}
\end{equation}
for the magnetization and the susceptibility, while the result for the
specific heat coincides with (\ref{eq:dissh}). Here, the critical
exponents keep their pure Ising values $\beta=1/8$ and $\gamma=7/4$,
but logarithmic corrections show up. Due to the slow variation of the
logarithm, it turns out to be very difficult to obtain reliable
results from numerical calculations, because the logarithmic
dependence on the reduced temperature translated to a logarithmic
dependence on the system size, and it is hopeless to distinguish a
simple logarithmic from a double logarithmic from finite-size
calculations. Still, the majority of the numerous finite-size scaling
studies based on Monte Carlo simulations or transfer matrix
calculations favour the latter result (\ref{eq:SSL}), see
\cite{Stinch,Selke,SST} and references therein, which was also
substantiated by a recent series expansion investigation \cite{RAJ}.

The correlation exponent of the three-dimensional pure Ising model
$\nu\approx 0.63$, thus $\alpha>0$, and disorder should be relevant in
that case. However, a recent result indicates that for purely
topological disorder introduced by considering an Ising model on a
three-dimensional random graph the critical behaviour stays the same
as for the regular cubic lattice \cite{JV}.

In recent years, there has been an increasing activity in the
mathematical literature on stochastic Ising model, see, e.g.,
\cite{New,BP,GHM}, leading to a number of rather general rigorous
results. For instance, for the random-field Ising model, it can be
shown rigorously that for dimensions $d\le 2$ there exists a unique
Gibbs state, thus there is \emph{no} phase transition in the
two-dimensional random-field model, whereas there is a phase
transition in $d\ge 3$. For ferromagnetic random-bond Ising models on
certain periodic planar graphs, it was proven that at most two
extremal Gibbs states exist, so there are no more than two phases, see
\cite{GH} and references therein.

\section{Aperiodic Ising Models}
\label{sec:aperiodic}

In the remainder of this article, we shall concentrate on the case of
aperiodically ordered Ising models \cite{GB}. To this end, we need to
introduce aperiodic sequences and tilings, see \cite{B} for a recent
review of the mathematical aspects. In the theoretical description of
quasicrystalline structures \cite{SSH}, aperiodic tilings constitute
the the analogues of periodic lattices in conventional
crystallography. For the sake of space, we cannot go into much detail
here. The interested reader is referred to the \emph{Mathematica}
\cite{wolfram} program packages accompanying this article, which give
an insight into the construction and properties of one- and
two-dimensional aperiodic tilings. The different methods employed and
the corresponding \emph{Mathematica} routines were described in detail
in \cite{GS}.

\subsection{Aperiodic Tilings on the Computer}

Aperiodic sequences are commonly constructed by means of
\emph{substitution rules}. As examples, consider an alphabet of two
letters $a$ and $b$ and rules
\begin{equation}
\varrho_{}^{(k)}:\;\;
\begin{array}{lcl}
a & \rightarrow & ab \\
b & \rightarrow & a^{k}_{}
\end{array}\qquad (k=1,2,\ldots)
\label{eq:subrule}
\end{equation}
that replace the single letters $a$ and $b$ by words $w_{a}^{(k)}=ab$
and $w_{b}^{(k)}=a^{k}_{}$, the word consisting of $k$ letters $a$, in
the two-letter alphabet $\{a,b\}$. We restrict ourselves to
\emph{primitive} substitutions, i.e., after a finite number of
iterations the words obtained from the basic letters should contain
all letters. Corresponding semi-infinite \emph{substitution sequences}
$w_{\infty}^{(k)}$ are then obtained by as fixed points of the
substitution rules by an iterated application $w_n^{(k)} =
\varrho_{}^{(k)}(w_{n-1}^{(k)})$ of the rules (\ref{eq:subrule}) on
some initial word, say $w_{0}^{(k)}=a$. The substitution rules
(\ref{eq:subrule}) yield some prominent examples of aperiodic
sequences, for instance the Fibonacci sequence $abaababa\ldots$ for
$k=1$, the so-called periodic-doubling sequence $abaaabab\ldots$ for
$k=2$, and a sequence known as the binary non-Pisot sequence
$abaaaababab\ldots$ for $k=3$.

Many properties of the substitution sequences can conveniently be
calculated form the associated substitution matrices
\begin{equation}
M_{}^{(k)} = \left(\begin{array}{@{\,}r@{\;\;}r@{\,}}
1&k\\ 1&0\end{array}\right)\; , \qquad
\lambda_{\pm}^{(k)} = \frac{1\pm\sqrt{4k+1}}{2}\; ,
\end{equation}
whose elements just count the number of letters $a$ and $b$ in their
substitutes $w_{a}^{(k)}$ and $w_{b}^{(k)}$, respectively. The largest
eigenvalue $\lambda_{+}^{(k)}$ of $M_{}^{(k)}$ determines the
asymptotic growth of the sequence in one iteration step, while the
entries of the corresponding eigenvector, suitably normalized,
determine the frequencies $p_{a}^{}=1-p_{b}^{}=1-2/(3+\sqrt{4k+1})$ of
the letters $a$ and $b$ in $w_{\infty}^{(k)}$. The second eigenvalue
measures the \emph{fluctuations} of the letter frequencies
\begin{equation}
g_{}^{(k)}(N) = \#_{a}[{w^{(k)}_{\infty}}_{|N}] - p_{a}^{} N \; , \qquad
h_{}^{(k)}(N) = 
\max_{M\le N}\left|g_{}^{(k)}(M)\right| 
\sim N_{}^{\omega^{(k)}_{}}\; ,
\end{equation}
where $\#_{a}[{w^{(k)}_{\infty}}_{|N}]$ denotes the number of letters
$a$ in the first $N$ letters of the limit word
$w^{(k)}_{\infty}$. Thus, $g_{}^{(k)}(N)$ measures te deviation from
the mean $p_{a}^{} N$, and we use the maximal deviation
$h_{}^{(k)}(N)$ to define the fluctuation exponent $\omega^{(k)}_{}$
which enters the Harris-Luck criterion, compare (\ref{eq:omega}).
It is given by 
\begin{equation}
\omega^{(k)}_{}=\frac{\ln|\lambda_{-}^{(k)}|}{\ln\lambda_{+}^{(k)}}\; ;
\qquad
\omega^{(1)}_{}=-1\; ,\quad
\omega^{(2)}_{}=0\; ,\quad
\omega^{(3)}_{}\approx 0.317>0\; ;
\end{equation}
and thus, according to the Harris-Luck criterion, a layered
aperiodicity introduced into the Ising model according to these
substitution rules is \emph{irrelevant} for $k=1$, \emph{marginal} for
$k=2$, and \emph{relevant} for $k=3$. So, contrary to what one might
have expected, \emph{deterministic} substitution sequences, albeit
being non-random, actually provide examples for all interesting
classes of `disorder'. 

\begin{figure}[tb]
\begin{center}
\includegraphics[width=0.69\textwidth]{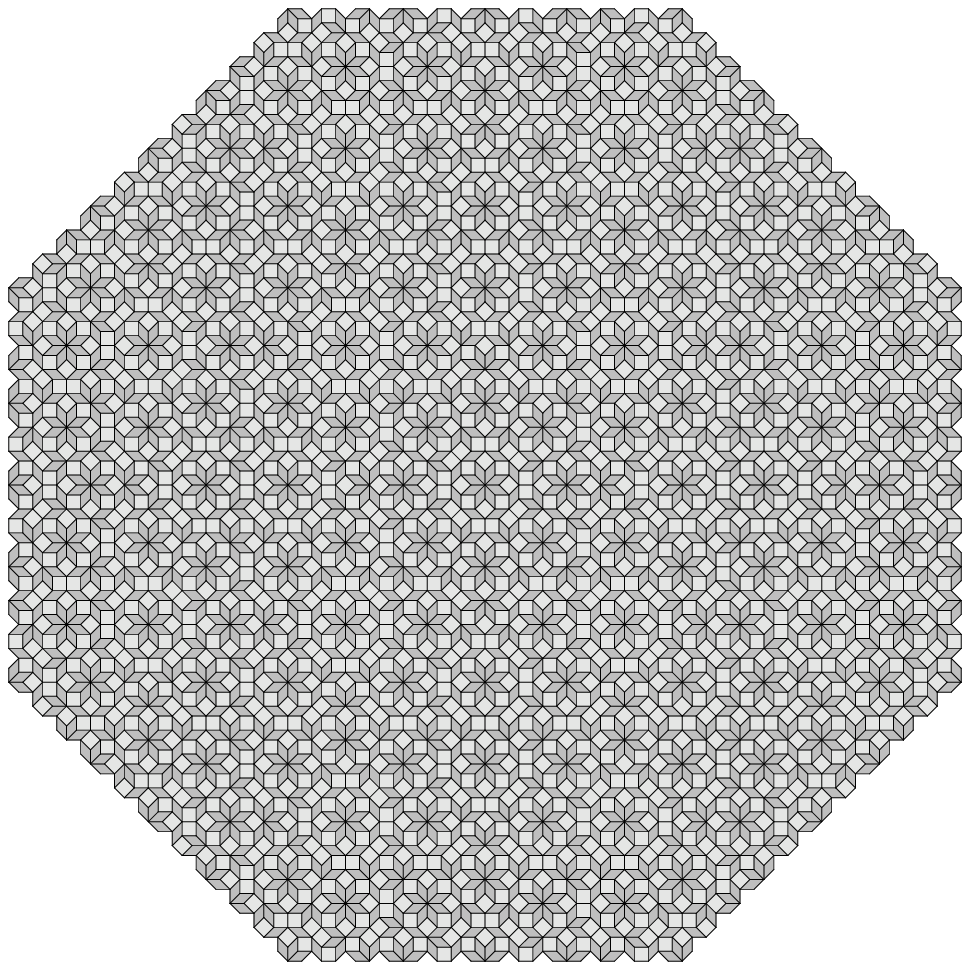}
\end{center}
\caption[]{Eightfold patch of the Ammann-Beenker tiling}
\label{fig:oct}
\end{figure}

The \emph{Mathematica} program {\tt FibonacciChain.m} that is included
in the program package offers the reader the opportunity to gain
experience with substitution sequences and their properties. In
addition, it also shows how quasiperiodic examples, such as the
Fibonacci chain, can be derived from a \emph{projection} of the
two-dimensional square lattice. This approach, known as the
cut-and-project method, can be generalized to higher dimensions to
obtain quasiperiodic tilings (or in mathematical terminology, model
sets \cite{B}) with interesting symmetry properties. Roughly, a
certain part of the high-dimensional periodic lattice is projected
onto a suitably chosen subspace. As an example, Fig.~\ref{fig:oct}
shows a central patch of an eightfold symmetric planar tiling obtained
from projection of the hypercubic lattice $\mathbb{Z}^4$. The tiling
consists of squares and rhombi and is known as the octagonal or the
Ammann-Beenker tiling \cite{ABS}. In addition, this geometric
structure also admits an inflation/deflation symmetry that is
analogous to the substitution rule for the one-dimensional
sequences. It consists of a dissection of the two basic tiles into
copies of itself, such that the resulting tiling is, apart from an
overall scaling, invariant under the procedure.

Both approaches are used in the \emph{Mathematica} program {\tt
OctagonalTiling.m} \cite{GS}. The remaining programs {\tt
ChairTiling.m} and {\tt SphinxTiling.m} deal with two further planar
examples of inflation tilings, while {\tt GridMethod.m} introduces a
variant of the cut-and-project scheme \cite{GS}. The package {\tt
PenrosePuzzle.m} employs yet another method to construct the most
famous among the zoo of quasiperiodic tilings, Penrose's tiling of the
plane. Here, the two basic rhombi are marked by certain arrow
decorations on their edges, and it is the task to fill the entire
plane with tiles without holes or overlaps, and without violating the
\emph{matching rules} that require that arrow decorations on adjacent
edges have to match.  The reader should be warned that this, albeit
seemingly simplest, puzzle method is not suited to construct a large
patch of the tiling because the matching rules do not uniquely specify
how to add tiles, and, in particular, do not prevent implicit mistakes
which will only be revealed when, at a certain stage, no further legal
additions of tiles are possible. But, if you indeed managed to fill
the entire plans without violating the rules, then the resulting
tiling is a Penrose tiling with all the other `magic' properties such
as an inflation/deflation symmetry.

\subsection{Ising Models on Planar Aperiodic Graphs}

In what follows, we shall consider planar Ising models defined on
quasiperiodic graphs, and, subsequently, Ising quantum chains with
coupling constants varying according to substitution sequences. The
latter correspond to two-dimensional classical Ising models with a
layered aperiodicity, thus $d_{\mathrm{m}}^{}=1$ in Harris-Luck
criterion, and $\omega_{\mathrm{c}}^{}=0$.

It may be surprising, but it is indeed possible to construct planar
aperiodic Ising models that are exactly solvable \cite{BGB,GB} in the
sense of commuting transfer matrices \cite{Bax}.  For such systems,
the coupling constants have to be chosen in a particular way to ensure
the integrability \cite{BGB}. As a consequence, they have the rather
unusual property that the free energy does \emph{not} depend on the
actual distribution of coupling constant, but only on the frequencies
of the various couplings. Another evidence of the particular
arrangement is the fact that the local magnetization, i.e., the
expectation value of a spin at a certain site, does not depend on the
site, but is uniform on the entire system. Therefore, the geometric
arrangement of couplings does not play a r\^{o}le, and the critical
behaviour is always the same as in the pure system.  Nevertheless,
these models are interesting because they provide counter-examples to
the Harris-Luck criterion, because the previous statement also holds
for arrangements with strong fluctuations that should have been
relevant. However, the solvable examples are certainly very special,
and possess some hidden underlying symmetry; and one should expect
that the criterion holds for any \emph{generic} distribution of
coupling constants.  Still, it reveals the limited predictive power of
such criteria.

Ising models on quasiperiodic graphs, in particular the Penrose
tiling, have been thoroughly investigated by means of Monte Carlo
simulations and by approximative renormalization group treatments, see
\cite{GB} and the literature cited therein. Usually, the coupling is
taken uniform along the bonds of the graph, so the fluctuations arise
solely from the locally differing coordination numbers. This type of
planar aperiodicity is irrelevant according to the Harris-Luck
criterion, because the fluctuations for quasiperiodic cut-and-project
sets, such as the Penrose tiling or the Ammann-Beenker tiling shown in
Fig.~\ref{fig:oct}, are small. Essentially, this is due to the fact
that the quasiperiodic tiling is the projection of a slab of a
periodic lattice, and thus, despite being aperiodic, these tilings
show a pronounced regularity. The numerical results unanimously
corroborate the prediction of the Harris-Luck criterion, and,
undoubtedly, these models belong to the same universality class as the
square-lattice Ising model.

Apart from these predominantly numerical approaches, also some
analytical techniques have been employed. This concerns, for
instance, high- and low-temperature expansions \cite{Domb} that can
be adapted to quasiperiodic systems.  For the Penrose and the
Ammann-Beenker tiling, the high-temperature expansion of the relevant 
part $\tilde{f}$ of the free energy,
\begin{equation}
\tilde{f} = \sum_{n=2}^{\infty}\, g_{2n}^{}\, w_{}^{2n} \; , \qquad
w = \tanh(\beta J)\; ,
\label{eq:hte}
\end{equation}
where $J$ denotes the coupling constant, has recently been calculated
to the 18th order in $w$ \cite{RGS1}. The coefficients $g_{2n}^{}$ can
be calculated from the frequencies of certain graphs of circumference
$2n$ in the tiling, which, for cut-and-project sets, can be calculated
explicitly.  Furthermore, each graph carries a certain weight; these
weights can be computed recursively. As an example, the expansion for
the Ammann-Beenker tiling is given by
\begin{eqnarray*}
&&\textstyle
w^{4} + \lambda w^{6} + (47\frac{1}{2}\! -\! 17\lambda) w^{8} +
(138\! -\! 50\lambda) w^{10} + 
(803\frac{1}{3}\! -\! 310\frac{1}{2}\lambda) w^{12}\\
&&\textstyle
+ (586\lambda\!-\! 1220) w^{14}
+ (96\frac{3}{4}\! +\! 295\frac{1}{2}\lambda) w^{16}
+ (46566\frac{1}{3}\lambda\!- \!108706) w^{18} + O(w^{20})
\end{eqnarray*}
where $\lambda=1+\sqrt{2}$. This may be compared with
the square-lattice result \cite{Domb}
\[
\textstyle
w^{4} + 2 w^{6} + 4\frac{1}{2} w^{8} + 
12 w^{10} + 37\frac{1}{3} w^{12} + 130 w^{14} + 
490\frac{1}{4} w^{16} + 1958\frac{2}{3} w^{18} + O(w^{20}) 
\]
where the coefficients are rational numbers. {}From the radius of
convergence of the series, and the behaviour close to it, one can, in
principle, derive the critical temperature and the thermal critical
exponent.  However, the information about the critical behaviour that
one can extract from the expansion is rather poor, because, in
contrast to the square-lattice case, the extrapolated values show
extremely strong fluctuations \cite{RGS1,RGS2}, and much more terms
would be necessary in order to give reliable estimates of the critical
temperature and the critical exponents. But this is not feasible; a
total of $244\, 638$ different subgraphs of the Ammann-Beenker tiling
contribute to the coefficient $g_{18}$, as compared with a mere $1975$
different graphs in the square-lattice case. A series analysis of
other quantities such as the magnetic susceptibility may yield more
stringent evidence, but this has not yet been calculated because the
number of graphs that have to be considered is even larger.

\begin{figure}[tb]
\begin{center}
\includegraphics[width=0.6\textwidth]{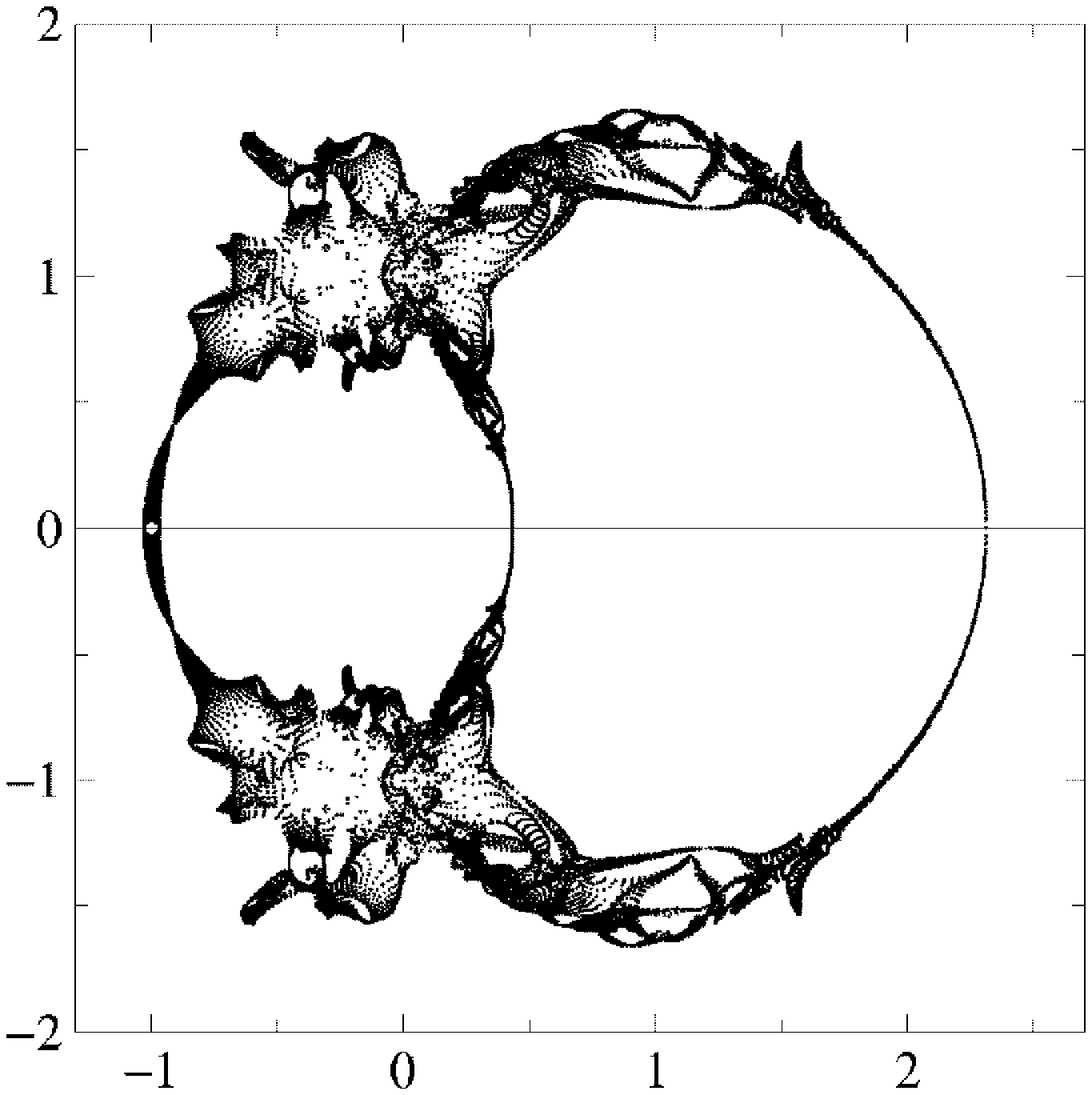}
\end{center}
\caption[]{Part of partition function zeros in the complex variable 
$z=\exp(2J/k_{\mathrm{B}}^{}T)$  for a periodic approximant of the 
Ammann-Beenker tiling with $41$ spins per unit cell}
\label{fig:zeros}
\end{figure}

Another method that relies on the computation of the zeros of the
partition function, in the complex temperature variable
$z=\exp(2J/k_{\mathrm{B}}^{}T)$. The set of zeros accumulates on
certain curves or areas in the complex plane that separate different
analytic domains, hence different phases of the system. Thus, a zero
on the real positive axis will correspond to a phase transition point
of the model, and the zeros close to this point contain information
about the critical exponents. For periodic approximants of aperiodic
tilings with rather large unit cells, the zero patterns can be
calculated explicitly \cite{RGS2}. The result for an approximant of
the Ammann-Beenker tiling, shown in Fig.~\ref{fig:zeros}, is rather
involved, whereas for the square lattice case, the zeros are
restricted to two circles with radius $\sqrt{2}$, centred at $z =\pm
1$. The two intersections of the zero pattern with the real axis
correspond to the ferromagnetic ($J>0$) and antiferromagnetic ($J<0$)
critical point, respectively, which are related to each other because
the tiling is bipartite. The corresponding estimates of the critical
temperature are given in Table~\ref{tab:AB}. Clearly, this method
allows a very precise determination of the critical temperature, 
in good agreement with, less accurate, Monte Carlo results \cite{Oli}.

\begin{table}[b]
\caption{Critical temperatures $w_{\mathrm{c}}^{}=\tanh
(J/K_{\mathrm{B}}^{}T_{\mathrm{c}}^{})$ for periodic approximants of
the Ammann-Beenker tiling, and estimated value for the aperiodic
tiling\label{tab:AB}}
\begin{center}
\begin{tabular*}{0.6\columnwidth}{c@{\extracolsep{\fill}}r@{\extracolsep{\fill}}l}
\hline
\rule[-1.5ex]{0ex}{4.5ex}%
$m$ &
\multicolumn{1}{c}{$N$} &
\multicolumn{1}{c}{$w_{c}$} \\ \hline
\rule[0ex]{0ex}{3.25ex}%
$1$      &     $7$  & $0.396\, 850\, 570$ \\
$2$      &    $41$  & $0.396\, 003\, 524$ \\
$3$      &   $239$  & $0.395\, 985\, 346$ \\
$4$      &  $1393$  & $0.395\, 984\, 811$ \\
$5$      &  $8119$  & $0.395\, 984\, 795$ \\
\rule[-1.5ex]{0ex}{1.5ex}%
$6$      & $47321$  & $0.395\, 984\, 795$ \\ \hline
\rule[-1.5ex]{0ex}{4.25ex}%
$\infty$ & $\infty$ & $0.395\, 984\, 79(1)$ \\ \hline
\end{tabular*}
\end{center}
\end{table}

\subsection{Aperiodic Ising Quantum Chains}

We now turn our attention to the case of one-dimensional aperiodicity
and consider Ising quantum chains with aperiodically modulated
coupling constants \cite{GB}
\begin{equation}
H =-\frac{1}{2}
\left(\sum\limits_{j=1}^{N} 
     \varepsilon_{j}^{}\,\sigma^{x}_{j}\sigma^{x}_{j+1} + 
     \sum\limits_{j=1}^{N} h_{j}^{}\sigma^{z}_{j}\right) \; ,
\label{eq:disham}
\end{equation}
where, in contrast to (\ref{eq:ham}), the couplings
$\varepsilon_{j}^{}\in\{\varepsilon_{a}^{},\varepsilon_{b}^{}\}$ and
the transversal fields $h_{j}^{}\in\{h_{a}^{},h_{b}^{}\}$ are chosen
according to the $j$th letter of an aperiodic substitution sequence.

Such models can be treated by an exact real-space renormalization
approach, exploiting the recursive structure that is already inherent
in the substitution sequence \cite{IT,ITKS,HGB,HG,Her}.  The basic idea
behind this approach is that a renormalization step \emph{reverses} a
substitution step, i.e., the Hamiltonian is transformed into a
Hamiltonian at the previous substitution level, which just differs
from the original Hamiltonian by renormalized values of the
parameters. The renormalization transformation then becomes a mapping
in the finite-dimensional parameter space of the Hamiltonian, and the
renormalization transformation is exact because no additional
parameters have to be introduced.  This is very interesting because it
provides one of the few examples with an exact renormalization
transformation for a non-trivial model, keeping in mind that the usual
decimation approach does not work even for the square-lattice Ising
model. Furthermore, it works for \emph{arbitrary} substitution
sequences \cite{HGB,Her}, not just for our examples
(\ref{eq:subrule}), and can even be applied to the case of random
substitutions \cite{Her}. This encompasses a large class of models
with different fluctuations in the couplings and, consequently,
different physical behaviour. For the sake of space, we cannot go into
detail here, but just briefly sketch the calculation.

The first step consists of a mapping of the Hamiltonian
(\ref{eq:disham}) onto a model of free fermions,
\begin{equation}
H = \sum_{j=1}^{N} \Lambda_{j}^{}\,\eta_{j}^{\dagger}\eta_{j}^{} \: +\: C
\; ,
\qquad
\{\eta_{j}^{},\eta_{\ell}^{}\} =
\{\eta_{j}^{\dagger},\eta_{\ell}^{\dagger}\} = 0 \; , \quad
\{\eta_{j}^{\dagger},\eta_{\ell}^{}\} = \delta_{j,\ell}^{}\; ,
\end{equation}
by means of a Jordan-Wigner transformation \cite{LSM}. In this way, the
task of computing the spectrum of (\ref{eq:disham}) is reduced from
the diagonalization of a $2^{N}_{}\times 2^{N}_{}$ matrix to that of
the $2N\times 2N$ eigenvalue problem
\begin{equation}
\left(
\begin{array}{@{}cccccccc@{}}
0&h_{1}^{}&0&0&0&\cdots&0&\varepsilon_{N}^{}\\
h_{1}^{}&0&\varepsilon_{1}^{}&0&0&\cdots&0&0\\
0&\varepsilon_{1}^{}&0&h_{2}^{}&0&\cdots&0&0\\
0&0&h_{2}^{}&0&\varepsilon_{2}^{}&\cdots&0&0\\
\vdots&\vdots&\ddots&\ddots&\ddots&\ddots&&\vdots\\
0&0&\cdots&0&\makebox[0pt]{$h_{\scriptscriptstyle N-1}^{}$}&
0&\makebox[0pt]{$\varepsilon_{\scriptscriptstyle N-1}^{}$}&0\\
0&0&\cdots&0&0&\makebox[0pt]{$\varepsilon_{\scriptscriptstyle N-1}^{}$}
 &0&h_{N}^{}\\
\varepsilon_{N}^{}&0&\cdots&0&0&0&h_{N}^{}&0
\end{array}\right)
\left(
\begin{array}{l@{}}
\Phi_{1}^{}\\
\Psi_{1}^{}\\
\Phi_{2}^{}\\
\Psi_{2}^{}\\
\;\vdots\\
\Psi_{\scriptscriptstyle N-1}^{}\\
\Phi_{N}^{}\\
\Psi_{N}^{}\\
\end{array}
\right) = \Lambda\left(
\begin{array}{l@{}}\Phi_{1}^{}\\
\Psi_{1}^{}\\
\Phi_{2}^{}\\
\Psi_{2}^{}\\
\;\vdots\\
\Psi_{\scriptscriptstyle N-1}^{}\\
\Phi_{N}^{}\\
\Psi_{N}^{}\\
\end{array}
\right)
\label{eq:mat}
\end{equation}
whose $N$ positive solutions
$0\le\Lambda_{1}^{}\le\Lambda_{2}^{}\le\ldots\le\Lambda_{N}^{}$
determine the entire spectrum of the Hamiltonian
(\ref{eq:disham}). The quantum phase transition appears when the
lowest excitation energy tends to zero, $\Lambda_{1}^{}\rightarrow 0$,
in the thermodynamic limit $N\rightarrow\infty$, so we can concentrate
on the low-energy behaviour.

The renormalization equations can now be derived by exploiting the
recursive structure of the sequence, effectively combining those sites
that belong to a single letter at the previous stage of the
substitution rule. This results in a mapping in parameter space, which
can be expanded in powers of $\Lambda$ at the fixed point $\Lambda=0$,
which corresponds to the critical point given by the condition
\begin{equation}
1=\lim_{N\rightarrow\infty} \left|
\frac{\varepsilon_{1}^{}}{h_{1}^{}}\frac{\varepsilon_{2}^{}}{h_{2}^{}}
\ldots\frac{\varepsilon_{N}^{}}{h_{N}^{}}\right|^{\frac{1}{N}} = 
\left|\frac{\varepsilon_{a}^{}}{h_{a}^{}}\right|^{p_{a}^{}}
\left|\frac{\varepsilon_{b}^{}}{h_{b}^{}}\right|^{p_{b}^{}}
\end{equation}
where the last equality applies to the two-letter case discussed
above. It can be parametrized as
$|\varepsilon_{a}^{}/h_{a}^{}|=r^{-p_b^{}}_{}$ and
$|\varepsilon_{b}^{}/h_{b}^{}|=r^{p_a^{}}_{}$ with
$r\in\mathbb{R}^{+}_{}$.  The mapping determines the finite-size
scaling of the smallest excitation energy $\Lambda_{1}$, and thus the
critical behaviour. In complete accordance with the Harris-Luck
criterion, we find that $\Lambda_{1}\sim v(r)N^{-1}$ for irrelevant
aperiodicity, as in the periodic case. For marginal aperiodicity, we
observe $\Lambda_{1}^{}\sim N^{-z(r)}$ with a coupling-dependent
non-universal exponent $z(r)$, corresponding to an anisotropic scaling
of the correlation length with different exponents
$\nu_{\parallel}^{}$ and $\nu_{\perp}^{}$. For relevant aperiodicity,
we find that the lowest excitation energy vanishes exponentially,
$\Lambda_{1}^{}\sim\exp[-\Delta(r)N^{\omega}]$, where $\omega$ is the
fluctuation exponent (\ref{eq:omega}). In fact, we can calculate also
the non-universal coupling-dependent terms. For our examples at hand,
this yields $v(r)=2\ln(r)/(r-1/r)$ for the Fibonacci chain $k=1$
(\ref{eq:subrule}), the exponent $z(r)=\ln(r^{1/3}+r^{-1/3})/\ln(2)$
for the period-doubling sequence $k=2$ (\ref{eq:subrule}), and the
coefficient $\Delta(r)=\sqrt{2\sqrt{13}-7}|\ln(r)|$ for the binary
non-Pisot sequence $k=3$ (\ref{eq:subrule}). This renormalization
approach can also be applied to obtain information about the
eigenvector in (\ref{eq:mat}), thus determining the behaviour of the
surface magnetization in an open quantum chain \cite{HG}.
Furthermore, other free-fermion models such as the XY chain can also
be treated \cite{Her,Joachim}.

\section{Summary and Conclusions}
\label{sec:conc}

The article gives a brief overview on recent results for aperiodic
Ising models. According to the personal taste and scientific
experience of the author, aperiodically ordered Ising models received
most attention, while random Ising models were only discussed from a
qualitative perspective. It is shown that heuristic scaling arguments
like the Harris-Luck criterion can provide a rather powerful tool
that, in most cases, correctly predict whether or not a certain
disorder will affect the critical behaviour. Still, behind the scene
are a number of tacit assumptions, and there exist at least
exceptional systems which defy the relevance criterion, as shown by an
exactly solvable example.

It is quite a difficult task to treat randomly disordered or
aperiodically ordered planar Ising models analytically, and rather
sophisticated techniques have been employed. However, it turns out
that the Ising quantum chain with coupling constants modulated
according to substitution sequences can be solved by an exact
real-space renormalization approach, thus proving the validity of the
Harris-Luck criterion for this entire class of models and, in addition,
providing information about the non-universal coefficient that enter
the scaling behaviour.

\section*{Acknowledgements}

The author thanks Michael Baake, Anton Bovier, Joachim Hermisson,
Wolfhard Janke, and Przemy{\l}aw Repetowicz for helpful
discussions.


\begin{thebibliography}{99.}
\addcontentsline{toc}{section}{References}

\bibitem{Bal}
R.~Balian:
\emph{From Microphysics to Macrophysics}, vol. I
(Springer, Berlin 1991)

\bibitem{Fre}
D.~Frenkel:
Physica A \textbf{263}, 26 (1999)

\bibitem{Bax}
R.J.~Baxter:
\emph{Exactly Solved Models in Statistical Mechanics}
(Academic Press, London 1982)

\bibitem{Lenz}
W.~Lenz:
Physikalische Zeitschrift \textbf{21}, 613 (1920)

\bibitem{Ising}
E.~Ising:
Zeitschrift f\"{u}r Physik \textbf{31}, 253 (1925)

\bibitem{Ons} 
L.~Onsager:
Phys.\ Rev.\ \textbf{65}, 117 (1944)

\bibitem{SSH}
J.-B.~Suck, M.~Schreiber, P.~H\"{a}ussler (eds.):
\emph{Quasicrystals}
(Springer, Berlin 2001) to appear

\bibitem{Kada}
L.P.~Kadanoff:
`Scaling, universality and operator algebras'.
In: \emph{Phase Transitions and Critical Phenomena} vol.~5a,
ed.\ by C.~Domb, J.L.~Lebowitz
(Academic Press, London 1976)
pp.~1--34

\bibitem{Binder}
K.~Binder (this volume)

\bibitem{RAR}
R.~R\"{o}mer (this volume)

\bibitem{Schwabl}
F.~Schwabl (this volume)

\bibitem{TV}
T.~Vojta (this volume)

\bibitem{Car}
J.L.~Cardy:
\emph{Scaling and Renormalization in Statistical Physics}
(Cambridge University Press, Cambridge 1996)

\bibitem{DFMS}
P.~DiFrancesco, P.~Mathieu, D.~S\'{e}n\'{e}chal:
\emph{Conformal Field Theory}
(Springer, New York 1997)

\bibitem{McCoyWu}
B.M.~McCoy, T.T.~Wu:
\emph{The two-dimensional Ising Model}
(Harvard University Press, Cambridge, Massachusetts, 1973)

\bibitem{Georgii}
H.-O.~Georgii:
\emph{Gibbs Measures and Phase Transitions} (de Gruyter, Berlin 1988)

\bibitem{Kobe}
S.~Kobe:
J.\ Stat.\ Phys.\ \textbf{88}, 991 (1997);
Phys.\ Bl\"{a}tter \textbf{54}, 917 (1998)

\bibitem{GB} 
U.~Grimm, M.~Baake:
`Aperiodic Ising models'. 
In: \emph{The Mathematics of Long-Range Aperiodic Order},
ed.\ by R.V.~Moody (Kluwer, Dordrecht 1997) pp.~199-237

\bibitem{Peierls}
R.~Peierls:
Proc.\ Cambridge Philos.\ Soc.\ \textbf{32}, 477 (1936)

\bibitem{Ellis}
R.S.~Ellis:
\emph{Entropy, Large Deviations, and Statistical Mechanics}
(Springer, New York 1985)

\bibitem{KW}
H.A.~Kramers, G.H.~Wannier:
Phys.\ Rev.\ \textbf{60}, 252 (1941)

\bibitem{Yang}
C.N.~Yang:
Phys.\ Rev.\ \textbf{85}, 808 (1952)

\bibitem{St}
R.B.~Stinchcombe:
%Ising model in a transverse field: I.~Basic theory,
J.\ Phys.\ C \textbf{6}, 2459 (1973)

\bibitem{LSM}
E.~Lieb, T.~Schultz, D.~Mattis: 
Ann.\ Phys.\ (NY) \textbf{16}, 407 (1961)

\bibitem{Kogut}
J.B.~Kogut:
Rev.\ Mod.\ Phys.\ \textbf{51}, 659 (1979) 

\bibitem{CBB}
C.~Chatelain, P.E.~Berche, B.~Berche:
Eur.\ Phys.\ J.\ B \textbf{7}, 439 (1999)

\bibitem{Her}
J.~Hermisson:
\emph{Aperiodische Ordnung und Magnetische Phasen\"{u}berg\"{a}nge}
(Shaker, Aachen 1999)

\bibitem{Har}
A.B.~Harris:
J.\ Phys.\ C:\ Solid State Phys.\ \textbf{7}, 1671 (1974)

\bibitem{Luck}
J.M.~Luck:
J.\ Stat.\ Phys.\ \textbf{72}, 417 (1993);
Europhys.\ Lett.\ \textbf{24}, 359 (1993)

\bibitem{Engel}
A.~Engel (this volume)

\bibitem{Stinch}
R.B.~Stinchcombe:
`Dilute magnetism'.
In: \emph{Phase Transitions and Critical Phenomena} vol.~7,
ed.\ by C.~Domb, J.L.~Lebowitz
(Academic Press, London 1983)
pp.~151--280

\bibitem{Selke}
W.~Selke,
`Monte Carlo simulations of dilute Ising models'.
In: \emph{Computer Simulation Studies in Condensed Matter Physics IV},
ed.\ by D.P.~Landau, K.K.~Mon, H.-B.~Sch\"{u}ttler
(Springer, Berlin 1993) pp.~18--27

\bibitem{SST}
W.~Selke, L.N.~Shchur, A.L.~Talapov:
`Monte Carlo simulations of dilute Ising models'.
In: \emph{Annual Reviews of Computational Physics I},
ed.\ by D. Stauffer 
(World Scientific, Singapore 1994) pp.~17--54

\bibitem{DD}
Vik.~S.~Dotsenko, Vl.~S.~Dotsenko:
Adv.\ Phys.\ \textbf{32}, 129 (1983)

\bibitem{Shalaev}
B.N.~Shalaev: Sov.\ Phys.\ Solid State \textbf{26}, 1811 (1984);
Phys.\ Rep.\ \textbf{237}, 129 (194)

\bibitem{Shankar}
R.~Shankar: Phys.\ Rev.\ Lett.\ \textbf{58}, 2466 (1987);
\textbf{61}, 2390 (1988)

\bibitem{Ludwig}
A.W.W.~Ludwig: Phys.\ Rev.\ Lett.\ \textbf{61}, 2388 (1988);
Nucl.\ Phys.\ B \textbf{330}, 639 (1990)

\bibitem{RAJ}
A.~Roder, J.~Adler, W.~Janke:
Phys.\ Rev.\ Lett.\ \textbf{80}, 4697 (1998)

\bibitem{JV}
W.~Janke, R.~Villanova,
`Ising spins on 3D random lattices'.
In: \emph{Computer Simulation Studies in Condensed Matter Physics XI},
ed.\ by D.P.~Landau, H.-B.~Sch\"{u}ttler
(Springer, Berlin 1999) pp.~22--26

\bibitem{New}
C.M.~Newman:
\emph{Topics in Disordered Systems}
(Birkh\"{a}user, Basel 1997)

\bibitem{BP}
A.~Bovier, P.~Picco (eds.):
\emph{Mathematical Aspects of Spin Glasses and Neural Networks},
Progress in Probability 41 (Birkh\"{a}user, Boston 1998)

\bibitem{GHM}
H.-O.~Georgii, O.~H\"{a}ggstr\"{o}m, C.~Maes:
`The random geometry of equilibrium phases'.
Preprint math.PR/9905031,
to appear in:
\emph{Phase Transitions and Critical Phenomena},
ed.\ by C.~Domb, J.L.~Lebowitz 
(Academic Press, London)

\bibitem{GH}
H.-O.~Georgii, Y.~Higuchi:
J.\ Math.\ Phys.\ \textbf{41}, 1153 (2000)

\bibitem{B}
M.~Baake:
`A guide to mathematical quasicrystals'.
Preprint math-ph/9901014,
to appear in \cite{SSH}

\bibitem{wolfram}
S.~Wolfram: 
\emph{Mathematica: A System for Doing Mathematics by Computer},
2nd ed.\
(Addison-Wesley, Reading, Massachusetts 1991)

\bibitem{GS}
U.~Grimm, M.~Schreiber:
`Aperiodic tilings on the computer'.
Preprint cond-mat/9903010,
to appear in \cite{SSH}

\bibitem{ABS}
R.~Ammann, B.~Gr\"{u}nbaum, G.~Shephard:
Discrete Comput.\ Geom.\ \textbf{8}, 1 (1992)

\bibitem{BGB}
M.~Baake, U.~Grimm, R.J.~Baxter:
Int.\ J.\ Mod.\ Phys.\ B \textbf{8} 3579 (1994)

\bibitem{Domb}
C.~Domb:
`Ising model'.
in: \emph{Phase Transitions and Critical Phenomena} vol.~3
ed.\ by C.~Domb, M.S.~Green
(Academic Press, London 1974)
pp.~357--458

\bibitem{RGS1}
P.~Repetowicz, U.~Grimm, M.~Schreiber:
J.\ Phys.\ A:\ Math.\ Gen.\ \textbf{32}, 4397 (1999)

\bibitem{RGS2}
P.~Repetowicz, U.~Grimm, M.~Schreiber:
`Planar quasiperiodic Ising models'.
Preprint cond-mat/9908088

\bibitem{Oli}
O.~Redner, M.~Baake: J.\ Phys.\ A:\ Math.\ Gen.\ \textbf{33},
3097 (2000)

\bibitem{IT}
F.~Igl\'{o}i, L.~Turban:
Phys.\ Rev.\ Lett.\ \textbf{77}, 1206 (1996)

\bibitem{ITKS}
F.~Igl\'{o}i, L.~Turban, D.~Karevski, F.~Szalma:
Phys.\ Rev.\ B \textbf{56}, 11031 (1997)

\bibitem{HGB} 
J.~Hermisson, U.~Grimm, M.~Baake:
J.\ Phys.\ A:\ Math.\ Gen.\ \textbf{30}, 7315 (1997)

\bibitem{HG} 
J.~Hermisson, U.~Grimm:
Phys.\ Rev.\ B \textbf{57}, R673 (1998)

\bibitem{Joachim}
J.~Hermisson:
J.\ Phys.\ A:\ Math.\ Gen.\ \textbf{33}, 57 (2000)

\end{thebibliography}
\end{document}